\documentstyle[12pt,epsfig]{article}

\newcommand{\ba}{\begin{eqnarray}}
\newcommand{\ea}{\end{eqnarray}}
\begin{document}
\pagestyle{plain}
\title{Hypercentral constituent quark model and isospin dependence}

\author{M. M. Giannini, E. Santopinto and A. Vassallo\\}

\maketitle
\noindent
\begin{center}
Dipartimento di Fisica dell'Universit\`a di Genova,\\
I.N.F.N., Sezione di Genova\\
via Dodecaneso 33, 16146 Genova, Italy\\
e-mail:giannini@genova.infn.it\\
\end{center}
\vspace{4pt}
\begin{abstract}
The constituent quark model based on a hypercentral
approach takes into account three-body force effects and standard two-body
potential contributions. The quark potential contains a hypercentral
interaction, to which a hyperfine term is added. While the
hypercentral potential supplies good 
values for the centroid energies of the resonance multiplets and a
realistic set of quark wave functions, the hyperfine splittings are
sometimes not  sufficient to account for the observed masses.
 In this work we have introduced an
improved form of the hyperfine interaction and an isospin dependent
quark potential. The resulting description of the baryon spectrum is very good,
also for the Roper resonance, specially thanks to the flavour dependent
interaction.
\end{abstract}
\begin{center}
PACS numbers: 12.39.Jh,12.39Pn,14.20.Gk
\end{center}

\section{Introduction}
Constituent Quark Models have been recently widely used for the description 
of the internal structure of baryons \cite{is,ci,gi,olof,bil,pl}. The baryon 
spectrum is usually described well, although the various models are quite 
different. However the study of 
hadron spectroscopy is not 
sufficient to distinguish among the various forms of quark dynamics. To 
this end one has to study in a consistent way all the physical observables 
of interest, in particular, besides the spectrum, the photocouplings,
the electromagnetic form factors and the strong decay amplitudes.
Such a systematic study of baryon properties is better performed 
within a general framework, and in this respect a hypercentral 
approach to quark dynamics can be used \cite{pl}. 
 The model consists of a hypercentral quark interaction containing a linear 
plus coulomb-like term, as suggested 
by lattice QCD calculations \cite{lat,bali}. A hyperfine term of the standard 
form \cite{is} is added and treated as a perturbation. 
The few free parameters of the model are fitted to the spectrum, 
the resulting baryon states are then used in order to calculate the 
various properties of interest, in particular the photocouplings \cite{aie}, 
the transition form factors \cite{aie2,mds2} and the elastic electromagnetic 
form 
factors of the nucleon \cite{mds}. The electromagnetic properties are 
evaluated using a non relativistic current for pointlike quarks, also 
taking into account the effects of relativistic 
corrections \cite{mds2,mds}. In particular this parameter-free calculation 
predicts that the ratio of the electric and magnetic proton factors 
decreases with $Q^{2}$ \cite{rap}, as shown by the recent TJNAF experiment 
\cite{ped}.

The description of the non strange baryon spectrum obtained by the 
hypercentral Constituent Quark Model (hCQM)  \cite{pl} 
is fairly good and comparable 
to the results of other approaches. In particular, the SU(6)-structure of 
the levels is accounted for thanks to the spin-independent hypercentral 
interaction; the $\Delta-N$ mass 
difference is correctly described by the hyperfine splitting and the 
theoretical energies of the negative parity resonances are in good 
agreement with data. However, notwithstanding such overall fair description 
of the spectrum, in some cases the splittings within the various SU(6)-
multiplets are too low. This is particularly true for the Roper resonances 
and for the higher states. A possible origin of these problems could be the 
(widespread) use of a $\delta-$like hyperfine interaction. To this end we 
have introduced different kinds of space smearings; the resulting hyperfine 
term becomes acceptable from the
theoretical point of view, but, as we 
shall show below, it does not improve the description of the spectrum.

A more important issue is the flavour dependence of the quark interaction. 
Actually, within the algebraic approach, the quark energy is written 
in terms of Casimir operators of symmetry groups which are relevant for 
the three-quark dynamics; in this respect it is quite natural to introduce 
an isospin dependent term, which turns out to be important for the 
description of the spectrum \cite{bil,gura}. On the other hand, in the chiral 
constituent quark model recently proposed \cite{olof,ple}, the splittings are
produced by the 
one-boson-exchange interaction between quarks and therefore a 
flavour-dependent potential arises, which seems to be important in order to 
describe the baryon spectrum, at least below $1.7~GeV$.

In the following, we shall show that in the hCQM a flavour dependent 
potential can be introduced \cite{vass} as a perturbative term leading 
to improved splittings within the $SU(6)$-multiplets. 
In  particular, in this way, the Roper 
resonance is reproduced quite well and the higher states acquire a much 
larger splitting, in agreement with data.

In Section $2$ we remind briefly the model and the main results in
the description of the spectrum and the electromagnetic excitation of
the baryon resonances. In Section $3$ we introduce in the hCQM a 
generalized SU(6)-breaking interaction treated as a perturbation and 
we show the results of the model compared with the 
experimental  spectrum. Finally, in Section $4$ there are some discussions
 and conclusions.

\section{The hypercentral model}

The internal quark motion is described by the Jacobi coordinates
$\mbox{\bf $\rho$}$ and $\mbox{\bf {$\lambda$}}$:
\begin{eqnarray}
\mbox{\bf $\rho$}~=~ \frac{1}{\sqrt{2}}
(\mbox{\bf r}_1 - \mbox{\bf r}_2) ~,
~~~~\mbox{\bf $\lambda$}~=~\frac{1}{\sqrt{6}}
(\mbox{\bf r}_1 + \mbox{\bf r}_2 - 2\mbox{\bf r}_3) ~,
\end{eqnarray}
\noindent or equivalently, $\rho$, $\Omega_{\rho}$, $\lambda$, 
$\Omega_{\lambda}$.  In order to describe the three-quark dynamics it is 
convenient to introduce the
hyperspherical coordinates, which are obtained substituting the absolute
values $\rho$ and $\lambda$ by
\begin{equation}
x=\sqrt{{\mbox{\bf $\rho$}}^2+{\mbox{\bf $\lambda$}}^2} ~~,~~ \quad
\xi=arctg(\frac{{\rho}}{{\lambda}}),
\end{equation}
where $x$ is the hyperradius and $\xi$ the hyperangle. In this way one can use
the hyperspherical harmonic formalism \cite{hh}.

In the hypercentral constituent quark model (hCQM), the quark potential,
$V$, is assumed to depend on the hyperradius $x$ only, that is to be 
hypercentral. Therefore, $V~=~V(x)$ is in general a three-body potential,
since the
hyperradius $x$ depends on the coordinates of all the three quarks. Since 
 the potential depends on $x$ only, in the three-quark wave 
function one can factor out the hyperangular part, 
which is given by the known hyperspherical harmonics 
\cite{hh}. The remaining hyperradial part of the wave function is 
determined by the hypercentral Schr\"{o}dinger equation:
\begin{equation}
[\frac{{d}^2}{dx^2}+\frac{5}{x}\frac{d}{dx}-\frac{\gamma(\gamma+4)}{x^2}]
{\psi}_{[\gamma]}(x)=-2m[E-V(x)]{\psi}_{[\gamma]}(x),
\end{equation}\label{rad}
\noindent where ${\psi}_{[\gamma]}(x)$ is the hypercentral wave function
and $\gamma$ is the
grand angular quantum number given by $\gamma~=~2v~+~l_{\rho}~+
~l_{\lambda}$; $l_{\rho}$ and $l_{\lambda}$ are the angular momenta
associated with the $\mbox{\bf $\rho$}$ and $\mbox{\bf $\lambda$}$ variables and $v$ is 
a non negative integer number.

There are at least two hypercentral potentials which lead to 
analytical solutions. First,
the h.o. potential, which has a two-body character, turns out to be exactly
hypercentral, since
\begin{equation}
\sum_{i<j}~\frac{1}{2}~k~(\mbox{\bf r}_i - \mbox{\bf r}_j)^2~=~\frac{3}{2}~k~x^2~=
~V_{h.o.}(x).
\end{equation}
The second one is the 'hypercoulomb' potential \cite{hca,hyp,br}
\begin{equation}
V_{hyc}(x)= -\frac{\tau}{x}.
\end{equation}\label{hcb}
\noindent This potential is not confining, however it has interesting
properties. In fact it leads to a power-law behaviour of the
proton form factor \cite{hyp} and of all the transition form 
factors \cite{sig,sig1}. Moreover it has an exact degeneracy between
the first $0^+$ excited state and the first $1^-$ states \cite{rich,hyp,br},
which can be respectively identified with the Roper resonance and
the negative parity resonances. This degeneracy seems to be in agreement with
phenomenology and is typical of an underlying O(7) symmetry \cite{br}.
This feature cannot be reproduced in
models with only two-body forces and/or harmonic oscillator bases since the
excited $L = 0$ state, having one more node, lies above the $L =1 $ state 
\cite{rich}.

The dynamic symmetry $O(7)$ of the hyperCoulomb problem
can be used to obtain the eigenvalues using purely algebraic
methods, similarly to what is done in the hydrogen atom case with the $O(4)$
symmetry.  In fact, the
hyperCoulomb Hamiltonian can be rewritten as \cite{br}
\begin{equation}
~~~H~~=~~-\frac{{{\tau}^2}m}{2~[C_{2}(O(7))+\frac{25}{4}]}
~~~~~,
\end{equation}\label{ho7}
\noindent where $C_{2}(O(7))$ is the quadratic Casimir invariant of $O(7)$,
and the energy eigenvalues became
\begin{equation}
~~~E~~=~~-\frac{{{\tau}^2}m}{2(n+5/2)^2}~~~~,
\end{equation}\label{en}
\noindent where n is a non negative integer \cite{br}.

As a confining hypercentral potential in our model we have assumed a  
form \cite{pl}
\begin{equation}\label{eq:pot}
V(x)= -\frac{\tau}{x}~+~\alpha x~~~~,
\end{equation}
\noindent 
that means a coulomb-like term plus a linear confining term as suggested by 
lattice QCD calculations \cite{lat,bali}.
In order to describe the splittings within the $SU(6)$-multiplets 
we introduce a hyperfine interaction of the standard form \cite{is} and we treat
it as a perturbation. 
Having fixed the quark mass $m$ to $1/3$ of the nucleon mass, the 
remaining three free parameters ($\tau$, $\alpha$ and the strength of the 
hyperfine interaction) are fitted to the spectrum. The strength of the
hyperfine 
interaction is determined by the $\Delta$ - Nucleon mass difference and 
the spectrum is described with $\tau=4.59$
and $\alpha=1.61~fm^{-2}$.  
Having fixed the parameters 
of the potential, the wave functions of the various 
resonances are 
completely determined and have been used for the calculation of the 
photocouplings \cite{aie}, the transition form factors to the 
negative parity resonances \cite{aie2}, the elastic form factors 
\cite{mds} and the ratio between the electric and magnetic form factors of 
the proton \cite{rap}. 

The resulting overall description of the experimental data is quite good, 
however there are some problems.
As far as the spectrum is concerned, the model fails to reproduce the Roper 
resonance and the higher levels. The evaluated size of the nucleon is too 
low, since the resulting r.m.s. radius is $0.46 fm$ and this affects
the description of the elastic form factors. The introduction of 
relativistic corrections substantially improves the calculated form 
factors \cite{mds2,rap} but it is not enough. The strength of the 
helicity amplitudes are underestimated, as it happens in all 
constituent quark models. 

There may be different reasons for such discrepancies. The 
splittings within the multiplets are not all adequately described by the 
hyperfine interaction. One should remind that the form we have assumed for 
the hyperfine interaction contains a $\delta$-like term, which is 
troublesome from the theoretical point of view. Furthermore, the splittings 
can be originated also by other terms, for instance isospin-dependent ones 
\cite{bil,olof}.

Another possible reason of the discrepancies is the fact 
that the model does not contain any explicit quark-pair creation mechanism,
 which is expected to be particularly important 
for the description of the eletromagnetic excitation \cite{iac,aie,aie2} 
but may lead to some 
residual effects also in the spectrum \cite{nat,paton,isgur}. 
In the latter case, the 
creation of quark-antiquark pairs could be the microscopic origin 
of an isospin 
dependent part of the potential. In this respect the 
constituent quark 
potential has to be considered as an effective potential in the three-quark 
subspace, taking into account implicitly the missing Fock-space 
configurations.

\section{The spin and isospin splittings}

The standard hyperfine interaction is used in order to reproduce the 
splittings within the $SU(6)-$multiplets. As mentioned above, it contains a 
$\delta$-like term which is an illegal operator. For this reason we
have modified it by introducing a smearing factor given by a gaussian 
function of the quark pair relative 
distance \cite{vass}:
\begin{equation}\label{eq:srho}
{H}_{S}=~A_{S}~\sum_{i<j}~\frac{1}{(\sqrt{\pi}\sigma_S)^{3}}~
e^{-\frac{r_{ij}^2}{\sigma_S^2}}~({\mbox{\bf s}}_{i}\cdot {\mbox{\bf s}}_{j})\mbox{~~,}
\end{equation}
\noindent where $\mbox{\bf s}_i$ is the spin operator of the i-th quark and $r_{ij}$
is the relative quark pair coordinate. The results for the spectrum 
are shown in Fig. 1. The fitted parameters are
$\alpha=1.58fm^{-2}$,
$\tau=4.98$,$~A_{S}~=38.4~fm^2$,$~\sigma_S~=~0.8 fm$. The correct 
limit for vanishing smearing is obtained, in the sense that the spectrum of 
Ref. \cite{pl} is reproduced. 

We have also tried a smearing depending on the hyperradius $x$ only 
of the form \cite{vass}:
\begin{equation}\label{eq:sx}
{H}_{Spin}=B_{S}~\left(\frac{1}{\Lambda_S}~e^{-\frac{x}{\Lambda_S}}
\right )~\sum_{i<j}({\mbox{\bf s}}_{i}\cdot {\mbox{\bf s}}_j)\mbox{~~,}
\end{equation}
\noindent where $\mbox{\bf s}_{i}$ is the spin operator of the i-th quark and
$x$ is the hyperradius. The fitted parameters are
$\alpha~=~1.49~fm^{-2}$,
$\tau~=~5.01$,$B_{S}~=~196.4~fm^{2}$,$~\Lambda_S~=~1.6 fm$.  There is an 
improvement for the higher states, however there is a too strong degeneracy 
of the levels and the good agreement for the negative parity states 
obtained in the hCQM$~$ \cite{pl} is lost. 
Therefore in the following we shall use only the hyperfine interaction 
with a two-body smearing.


As quoted in the previous sections, there are different motivations for the 
introduction of a flavour dependent 
term in the three-quark interaction.
The well known Guersey-Radicati mass formula \cite{gura}
contains a flavour dependent term, which is essential for the description
of the strange baryon spectrum:
\begin{eqnarray}
\hat{M}= M_{0}+a\hat{C}_2(SU_{SF}(6))+b\hat{C}_2(SU_{F}(3))\nonumber\\
+b'\hat{C}_2(SU_{I}(2))+b''\hat{C}_1(U_Y(1))\nonumber\\
+b'''\hat{C}_2(U_Y(1))+\hat{C}_2(SU_{S}(2))~,
\end{eqnarray}
\noindent where $M_0$ is fixed for any $SU(6)$-multiplet, 
$\hat{C}_2(SU(N))$ is the Casimir of SU(N), $Y$ is the
hypercharge, $I$ is the total isospin and $S$ the total spin. 
For non strange baryons this formula implies an isospin dependence.
 In the algebraic description of baryon
properties \cite{bil},
the space part of the mass operator is written in terms of the generators
of the U(7) group, while for the internal degrees of freedom the 
Guersey-Radicati mass formula \cite{gura} is used.
In the chiral Constituent Quark Model \cite{olof,ple}, the non 
confining part of the 
potential is provided by the interaction with the Goldstone bosons, 
giving rise to a spin- and isospin-dependent part, which is crucial for the 
description of the spectrum for energies lower than $1.7 GeV$.
It has been also pointed out quite recently that an isospin dependence of
the quark potential can be obtained by means of quark exchange \cite{dm}. 
More generally, one can expect that the quark-antiquark pair production
can lead to an effective quark interaction containing an isospin (or
flavour) dependent term. On the other hand, the fact that the
constituent quark model does not contain explicitly this mechanism is
may be the reason why the low $Q^2$
-behaviour of the electromagnetic transition form factors is not
reproduced \cite{iac,aie}.

With these motivations in mind, we have introduced isospin dependent terms 
in the hCQM hamiltonian.

To this end we have added two terms in the three-quark hamiltonian 
with the hyperfine interaction of Eq.(\ref{eq:srho}). 
The first one depends on the isospin only and has the form:
\begin{equation}\label{eq:taurho}
{H}_{\mathrm{I}}=
~A_{I} \sum_{i<j}\frac{1}{(\sqrt{\pi}\sigma_I)^3}~
e^{-\frac{{\bf r}^2_{ij}}{\sigma_I^2}}({\mbox{\bf t}}_i \cdot {\mbox{\bf t}}_j)
\mbox{~,}
\end{equation}\label{eq:st}
\noindent where $\mbox{\bf t}_i$ is the isospin operator of the i-th quark and 
 $r_{ij}$ is the relative quark pair coordinate.
The second one is a spin-isospin interaction, given by
\begin{equation}\label{si}
{H}_{\mathrm{SI}}=
A_{SI}~\sum_{i<j}\frac{1}{(\sqrt{\pi}\sigma_{SI})^3}~e^{-\frac{r^2_{ij}}
{\sigma^2_{SI}}}({\mbox{\bf s}}_i \cdot {\mbox{\bf s}}_j)({\mbox{\bf t}}_i \cdot {\mbox{\bf t}}_j)\mbox{~,}
\end{equation}
\noindent where $\mbox{\bf s}_i$ and $\mbox{\bf t}_i$ are respectively the spin and isospin
operators of the i-th quark and
 $r_{ij}$ is the relative quark pair coordinate. 
The complete interaction is then given by
\begin{equation}\label{tot}                                         
H_{int}~=~V(x) +H_{\mathrm{S}} +
H_{\mathrm{I}} +H_{\mathrm{SI}}~.
\end{equation}                                                         
The resulting spectrum for the 3*- and 4*- resonances is shown in Fig.2 
and in Table I. 
The $N-\Delta$ mass difference is no more due only to the 
hyperfine interaction. 
In fact, in this model its contribution is only about $35\%$, 
the remaining splitting comes from the
spin-isospin term, $(50\%)$, and from the isospin one, $(15\%)$. 

It should also be noted that the inversion between the Roper and the 
negative parity resonances is almost entirely due to the spin-isospin 
interaction, as stated in Ref. \cite{olof}.

The tensor term coming from the hyperfine interaction 
has been kept as well, however its contribution to the spectrum 
is negligible.

In Table II  and Table III we list all the remaining states
predicted by the model and when possible we give a temptative assignation
to one- and two-star states. We see that the number of
predicted states is higher than the presently observed ones, that is, 
as in other models, we have the problem of missing resonances. 
It is interesting to observe that while the quality of the reproduction 
of the spectrum for the 3*- and 4*- resonances is quite the same 
for the various Constituent Quark Models \cite{is,ci,bil,olof,ple}, 
there are different predictions for the other states, 
that means for 1*, 2*  and missing resonances. 

Recently in a three-channel multi-resonance amplitude analysis 
it has been found evidence for a third low-lying $P_{11}$ state at 
$1740\pm 11~MeV$ \cite{zagreb}. The first two
 $P_{11}$ states 
at $1439\pm 19~MeV$ and $1729\pm 16~MeV$ correspond to the N(1440) and 
N(1710) of the PDG \cite{pdg}. In the hCQM the first three $P_{11}$
 states are at 
$1463~MeV$, $1752~MeV$ and $1828~MeV$ 
respectively. A new analysis of kaon photoproduction data  
\cite{ben} has shown evidence for a third $D_{13}$ resonance at $1895~MeV$, 
which can be described by one of the states predicted by the present model 
(see Table II).

\section{Discussions and conclusions}

A considerable improvement in the description of the spectrum is obtained 
with an isospin dependent potential.
As quoted in the previous section, a possible motivation of the
isospin-dependent terms of the quark interaction  is given  by
quark-antiquark pair production mechanisms.

This kind of mechanisms 
have been studied within the string model in connection
with the quenching problem of quark models in 
the meson spectrum \cite{nat,paton,isgur}.
Their contributions turn out to be mainly spin-independent and can 
be reabsorbed into the string parameter, so the residual part 
can be considered as a perturbative correction. 
This can be considered a suggestion also for the baryon case, 
even if there are differences. Therefore, any interaction 
leading to unperturbed states with reasonable
spin-averaged values of the energy levels can be considered to contain
implicitly this kind of
contribution. The splittings, which are in general spin and isospin dependent,
 can be treated perturbatively. 

In this article we have shown that the complete interaction 
including spin and isospin terms (see Eq.(14)$~$), reproduces the position of
the two Roper resonances of the nucleon, while keeping the good
description of the negative parity resonances. It should be noted
that also the higher states are accounted for.
The hypercentral potential is a good starting point for the 
construction of an unperturbed spectrum and leads to realistic quark states, as it is 
shown also by the reproduction of the e.m. form factors and transition 
form factors, which are sensitive to the wave functions. 
The configuration mixing 
is usually a higher order correction, apart from some special cases as 
the neutron form factor or the $S_{11}(1650)$ helicity amplitude etc. , for which the 
$SU(6)$-contribution is vanishing. In this respect, 
the forthcoming more precise data that will be soon available 
will supply valuable information concerning the possible
 $SU(6)$-breaking terms in the quark interaction.
Finally, one can observe that the various 
Constituent Quark Models give different 
results concerning the 1* and 2* states and the position and number of the 
missing resonances .  Therefore, the expected new data coming from the TJNAF will 
be very helpful in order to discriminate among thems.


\newpage

\begin{figure}
\caption{The spectrum obtained with the hypercentral potential, 
Eq.(2.8), and the spin dependent term, Eq.(3.1). The fitted parameters are
$\alpha=1.58fm^{-2}$,
$\tau~=~4.98$,$~A~=~38.4~fm^2$,$\sigma~=~0.8 fm$.}
\label{autonum}
\end{figure}

\begin{figure}
\caption{The spectrum obtained with the complete interaction of Eq.(3.6), 
that means the hypercentral potential, of Eq.(2.8), plus the
spin dependent term, the isospin interaction and the
spin-isospin one (Eqs.(3.1),(3.4),(3.5)).  The fitted parameters are
$\alpha=1.17 fm^{-2}$,
$\tau=4.95$,$A_{S}=67.4fm^{2}$,$\sigma_{S}=2.87fm$,
$A_{I}=51.7fm^{2}$,$\sigma_{I}=3.45fm$,
$A_{SI}=-106.2fm^{2}$,$\sigma_{SI}=2.31fm$.}
\label{complete}
\end{figure}


\begin{table}
\caption{ Mass Spectrum of Nonstrange Baryon Resonances.}
\catcode`?=\active \def?{\kern\digitwidth}
\begin{tabular}{lcccr}
Baryon & Status & $M_{exp}$ & $J^{\pi}$ & $M_{theor}$ \\ 
& & (MeV) & &(MeV)\\  
\hline
N(938)~$P_{11}$ & **** & 938 & $\frac{1}{2}^+$ &  $938$ \\ 
$\Delta (1232)~P_{33}$ & **** & 1230-1234 & $ \frac{3}{2}^+$ & $1232$ \\ 
N(1440)~$P_{11}$ & **** & 1430-1470 & $\frac{1}{2}^+$ &  $1463$ \\ 
$\Delta (1600)~P_{33}$ & *** & 1550-1700 & $\frac{3}{2}^+$ &  $1727$ \\ 
N(1535)~$S_{11}$ & **** & 1520-1555 & $\frac{1}{2}^-$ &  $1524$ \\ 
N(1520)~$D_{13}$ & **** & 1515-1530 & $\frac{3}{2}^-$ &  $1524$ \\ 
N(1650)~$S_{11}$ & **** & 1640-1680 & $\frac{1}{2}^-$ &  $1688$ \\ 
N(1700)~$D_{13}$ & *** & 1650-1750 & $\frac{3}{2}^-$ &  $1692$ \\ 
N(1675)~$D_{15}$ & **** & 1670-1785 & $\frac{5}{2}^-$ & $1668$ \\ 
$\Delta (1620)~S_{31}$ & **** & 1615-1675 & $\frac{1}{2}^-$ &  $1573$ \\ 
$\Delta (1700)~D_{33}$ & **** & 1670-1770 & $\frac{3}{2}^-$ &  $1573$ \\ 
N(1710)~$P_{11}$ & *** & 1680-1740 & $\frac{1}{2}^+$ & $1752$ \\ 
N(1720)~$P_{13}$ & **** & 1650-1750 & $\frac{3}{2}^+$ &  $1648$ \\ 
N(1680)~$F_{15}$ & **** & 1675-1690 & $\frac{5}{2}^+$ &  $1680$ \\ 
$\Delta (1910)~P_{31}$ & **** & 1870-1920 &  $\frac{1}{2}^+$ &  $1953$ \\ 
$\Delta (1920)~P_{33}$ & *** & 1900-1970 & $\frac{3}{2}^+$ & $1921$ \\ 
$\Delta (1905)~F_{35}$ & **** & 1870-1920 & $\frac{5}{2}^+$ &  $1901$ \\ 
$\Delta (1950)~F_{37}$ & **** & 1940-1960 & $\frac{7}{2}^+$ & $1955$ \\ 
$\Delta (1900)~S_{31}$ & *** & 1850-1950 & $\frac{1}{2}^-$ & $1910$ \\ 
\end{tabular}
\end{table}

\clearpage

\begin{table}
\catcode`?=\active \def?{\kern\digitwidth}

\caption{All calculated Nucleon Resonances (in MeV) below 2 GeV .
Tentative assignments of 1- and 2- star resonances are shown in brackets.}
\begin{tabular}{lcr}
State & M$_{theor}$ & Baryon \\
\hline
$N^{\frac{1}{2}^+}$ & $938$ & $N(938)~P_{11}$\\
$N^{\frac{1}{2}^+}$ & $1463$ & $N(1440)~P_{11}$\\
$N^{\frac{1}{2}^+}$ & $1752$ & $N(1710)~P_{11}$\\
$N^{\frac{1}{2}^+}$ & $1828$ & \\
$N^{\frac{1}{2}^+}$ & $1894$ & \\
$N^{\frac{1}{2}^+}$ & $1938$ &  $[N(2100)~P_{11}]$\\

$N^{\frac{1}{2}^-}$ & $1524$ & $N(1535)~S_{11}$\\
$N^{\frac{1}{2}^-}$ & $1688$ & $N(1650)~S_{11}$\\
$N^{\frac{1}{2}^-}$ & $1861$ & \\
$N^{\frac{1}{2}^-}$ & $2008$ &  $[N(2090)~S_{11}]$\\

$N^{\frac{3}{2}^-}$ & $1524$ & $N(1520)~D_{13}$\\
$N^{\frac{3}{2}^-}$ & $1692$ & $N(1700)~D_{13}$\\ 
$N^{\frac{3}{2}^-}$ & $1860$ & \\
$N^{\frac{3}{2}^-}$ & $2008$ &  $[N(2080)~D_{13}]$\\ 

$N^{\frac{3}{2}^+}$ & $1648$ & $N(1720)~P_{13}$\\
$N^{\frac{3}{2}^+}$ & $1816$ & \\
$N^{\frac{3}{2}^+}$ & $1894$ & $[N(1900)~P_{13}]$\\
$N^{\frac{3}{2}^+}$ & $1939$ & \\
$N^{\frac{3}{2}^+}$ & $2034$ & \\

$N^{\frac{5}{2}^+}$ & $1680$ & $N(1680)~F_{15}$\\
$N^{\frac{5}{2}^+}$ & $1833$ & \\
$N^{\frac{5}{2}^+}$ & $2046$ &  $[N(2000)~F_{15}]$\\

$N^{\frac{7}{2}^+}$ & $1939$ & $[N(1990)~F_{17}]$\\

$N^{\frac{5}{2}^-}$ & $1668$ & $N(1675)~D_{15}$\\
$N^{\frac{5}{2}^-}$ & $1984$ & $[N(2200)~D_{15}]$\\
\end{tabular}
\end{table}


\begin{table}
\catcode`?=\active \def?{\kern\digitwidth}
\caption{All calculated Delta Resonances (in MeV) below 2 GeV .
Tentative assignments of 1- and 2-star resonances are shown in brackets.}
\begin{tabular}{lcr}
State & M$_{theor}$ & Baryon \\
\hline
$\Delta^{\frac{1}{2}^+}$ & $1900$ & $\Delta(1910)~P_{31}$\\
$\Delta^{\frac{1}{2}^+}$ & $1953$ & \\

$\Delta^{\frac{1}{2}^-}$ & $1573$ & $\Delta(1620)~S_{31}$\\
$\Delta^{\frac{1}{2}^-}$ & $1910$ & $\Delta(1900)~S_{31}$\\

$\Delta^{\frac{3}{2}^+}$ & $1232$ & $\Delta(1232)~P_{33}$\\
$\Delta^{\frac{3}{2}^+}$ & $1727$ & $\Delta(1600)~P_{33}$\\
$\Delta^{\frac{3}{2}^+}$ & $1921$ & $\Delta(1920)~P_{33}$\\
$\Delta^{\frac{3}{2}^+}$ & $1955$ & \\
$\Delta^{\frac{3}{2}^+}$ & $2049$ & \\

$\Delta^{\frac{3}{2}^-}$ & $1573$ & $\Delta(1700)~D_{33}$\\
$\Delta^{\frac{3}{2}^-}$ & $1900$ & $[\Delta(1940)~D_{33}]$\\

$\Delta^{\frac{5}{2}^+}$ & $1901$ & $\Delta(1905)~F_{35}$\\
$\Delta^{\frac{5}{2}^+}$ & $1956$ & $[\Delta(2000)~F_{35}]$\\ 

$\Delta^{\frac{7}{2}^+}$ & $1955$ & $\Delta(1950)~F_{37}$ \\
\end{tabular}
\end{table}

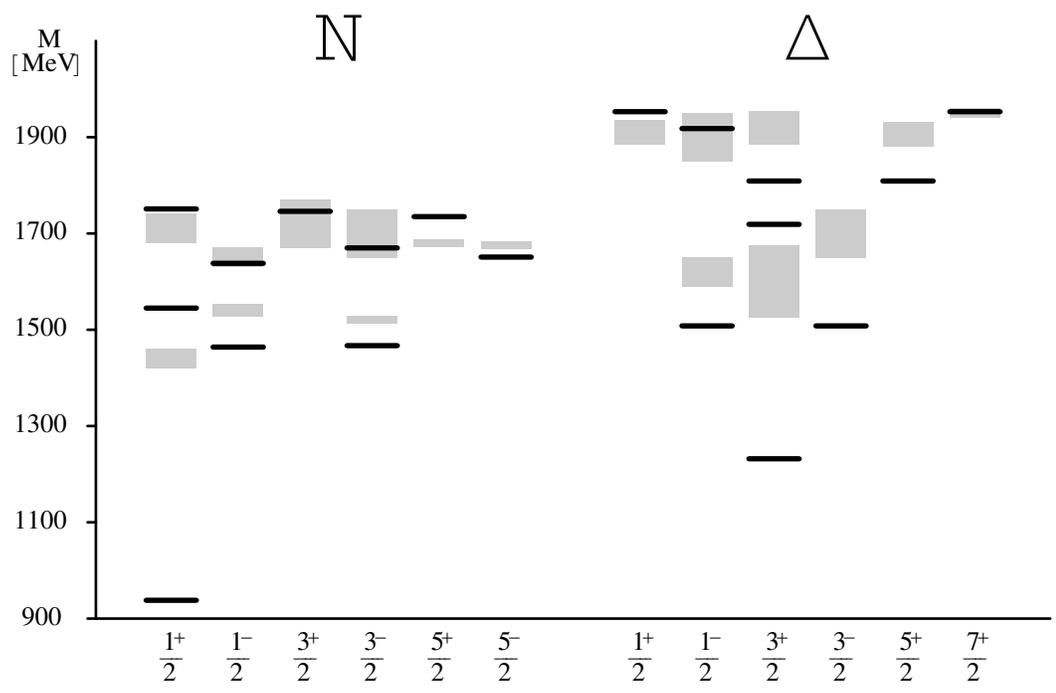
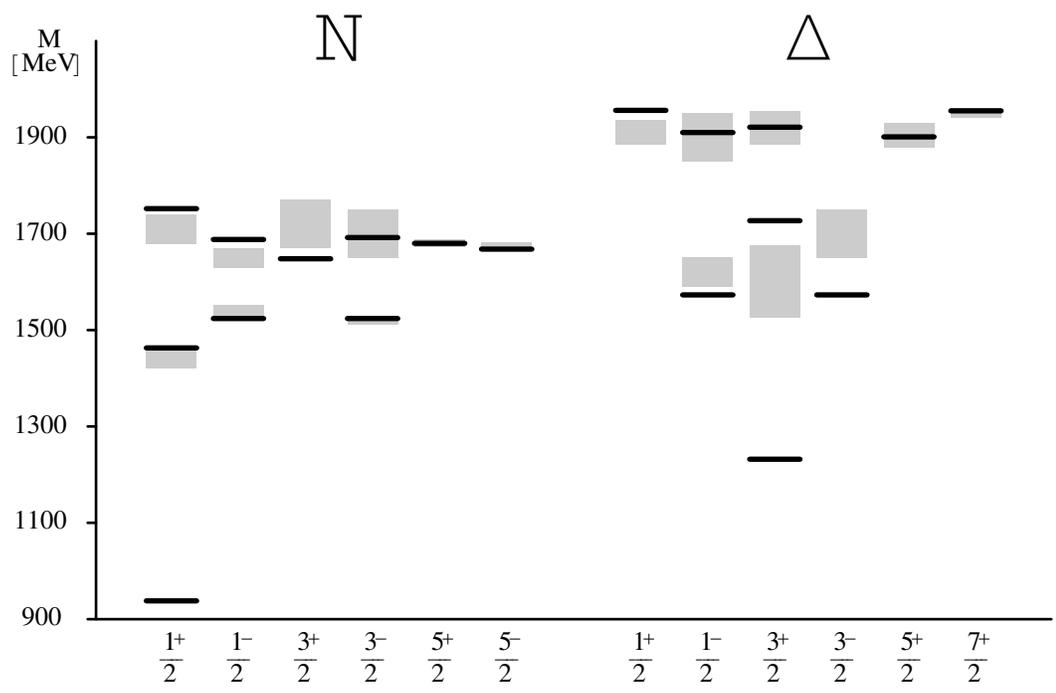
\end{document}